\documentclass[aps,prd,reprint,showpacs,preprintnumbers,superscriptaddress,nofootinbib]{revtex4-1}

\usepackage{slashed}
\usepackage{bm} 
\usepackage{epsf}
\usepackage{graphicx,epsfig}
\usepackage{latexsym,amssymb,amsmath,float,url}
\usepackage{bbold}
\usepackage{latexsym}
\usepackage{soul}

\input{epsf}

\begin{document}

\title{Baryon Number Violating Scalar Diquarks at the LHC}

\author{Iason\ Baldes}
\affiliation{ARC Centre of Excellence for Particle Physics at the Terascale,
School of Physics, The University of Melbourne, Victoria 3010, Australia}

\author{Nicole F.\ Bell} 
\affiliation{ARC Centre of Excellence for Particle Physics at the Terascale,
School of Physics, The University of Melbourne, Victoria 3010, Australia}

\author{Raymond R.\ Volkas}
\affiliation{ARC Centre of Excellence for Particle Physics at the Terascale,
School of Physics, The University of Melbourne, Victoria 3010, Australia}

\date{November 22, 2011}

\preprint{}

\begin{abstract}
Baryon number violating (BNV) processes are heavily constrained by experiments searching for nucleon decay and neutron-antineutron oscillations. If the baryon number violation occurs via the third generation quarks, however, we may be able to avoid the nucleon stability constraints, thus making such BNV interactions accessible at the LHC. In this paper we study a specific class of BNV extensions of the standard model (SM) involving diquark and leptoquark scalars. After an introduction to these models we study one promising extension in detail, being interested in particles with mass of $\mathcal{O}(\mathrm{TeV})$. We calculate limits on the masses and couplings from neutron-antineutron oscillations and dineutron decay for couplings to first and third generation quarks. We explore the possible consequences of such a model on the matter-antimatter asymmetry. We shall see that for models which break the global baryon minus lepton number symmetry, $(B-L)$, the most stringent constraints come from the need to preserve a matter-antimatter asymmetry. That is, the BNV interaction cannot be introduced if it would remove the matter-antimatter asymmetry independent of baryogenesis mechanism and temperature. Finally, we examine the phenomenology of such models at colliders such as the LHC.  
\end{abstract}



\maketitle

\section{Introduction}
Experiments so far have not detected any violation of baryon number ($B$), an accidental global symmetry of the standard model Lagrangian.  Its conservation is dramatically demonstrated by the stringent limits on the proton lifetime, roughly $\tau > 10^{31}$ years. The proton is not the lightest colour singlet, but it is the lightest colour singlet with baryon number, the conservation of which at low energy leads to its stability. 

Similarly stringent limits are set from searches of baryon number violating neutron decay modes, neutron-antineutron oscillations, and dinucleon decays. Baryon number violating processes must therefore be suppressed at low energy. The picture, however, may change at higher energy scales. There is no reason for $B$ to be conserved; on the contrary, there are many strong theoretical reasons why this global symmetry should be broken.

The SM already contains violation of baryon number at high temperatures, through the non-perturbative sphaleron process, which breaks $B$ and lepton number ($L$), but leaves ($B-L$) intact. There may exist copious other examples at high energy. Perhaps the most compelling theoretical reason for baryon number violation comes from the need to explain the matter-antimatter asymmetry. 

Observations of the cosmic microwave background and primordial deuterium abundance gives the ratio of baryon number density $n_{b}$, to photon number density $n_{\gamma}$ of \cite{Copi:1994ev,Bennett:2003bz}:
	\begin{equation}
	\frac{n_{b}}{n_{\gamma}}=(6.1\pm0.3)\times10^{-10}.
	\end{equation}
In a symmetric universe with $B=0$, the nucleons and antinucleons would continue annihilating until \cite{PhysRevLett.17.712,RevModPhys.53.1}:
	\begin{equation}
	\frac{n_{b}}{n_{\gamma}}=10^{-19}.
	\end{equation}
Assuming the universe starts off in a symmetric state, there must be a dynamical mechanism (baryogenesis), to create an asymmetry in baryon number. Baryogenesis presumably occurred at some high temperature, $T > 100\mathrm{GeV}$, and possibly much higher.

Sakharov listed the three required conditions: (i) baryon number violation, (ii) C and CP violation, and (iii) departure from thermal equilibrium \cite{Sakharov:1967dj}. Though there is some room for the non-perturbative sphaleron process to be the baryogenesis mechanism (electroweak baryogenesis) in SUSY scenarios, the parameter space is highly constrained. Alternative scenarios for baryogenesis have therefore also been proposed (for a review see \cite{Dine:2003ax,Chen:2007fv}).

Furthermore, baryon number violation may arise quite generically in new high energy physics. In this case, there exists not only the BNV process which gives rise to baryogenesis, but also other sources of BNV as well. Though not the mechanism for baryogenesis, these processes can affect the matter-antimatter asymmetry, even washing out an existing asymmetry \cite{PhysRevLett.45.2074}. We can use this to place additional constraints on new physics models or alternatively, if such additional BNV were detected, it may even be possible to rule out certain baryogenesis mechanisms as the source of the observed matter-antimatter asymmetry, as their contribution would simply be washed out by the additional BNV processes. 

Leptogenesis, for example, creates an asymmetry in the lepton sector at high temperature. This asymmetry clearly has a $(B-L)$ component. The $L$ asymmetry is then reprocessed into an asymmetry in the baryon sector by the rapid ($B+L$) violating, but $(B-L)$ conserving, sphaleron process. The introduction of additional BNV but ($B-L$) conserving interactions simply increases the rate of reprocessing. The addition of rapid ($B-L$) violating interactions, however, would lead to a washout of the entire asymmetry.

In this paper, we shall introduce baryon number violation by first augmenting the SM with scalar diquarks and leptoquarks. Introducing interactions between these scalars then leads to baryon number violation. Such interactions were originally motivated by charge quantization\cite{catalogue}, and have been studied as a possible mechanism for baryogenesis\cite{sigmabaryogenesis}. 

Since the diquarks and leptoquarks all carry electromagnetic and colour charge, the gauge interactions are too strong for the requisite departure from thermal equilibrium needed for baryogenesis to take place unless the new scalars have masses $\gg \mathcal{O}(\mathrm{TeV})$. In this paper we are interested in the phenomenology of these models at the LHC, that is with $\mathcal{O}(\mathrm{TeV})$ masses, so we will not be considering these interactions as a baryogenesis mechanism here. 

BNV interactions involving the first generation quarks are strongly constrained by nucleon stability. It is possible to use the nucleon stability results to put stringent limits on dimension six BNV operators (the lowest dimension effective BNV operators possible) involving the higher generations, meaning that BNV interactions are unlikely to be observed at the LHC \cite{Hou:2005iu}. If some unknown GIM-like cancellation mechanism were to suppress the nucleon instability, however, BNV interactions involving the top quark may become accessible at the LHC \cite{Dong:2011rh}.

Instead of searching for a GIM-like cancellation mechanism, by examining the flavour structure of new physics models, we concentrate on BNV that proceeds through multiple heavy scalars. Each additional scalar of mass $M$ suppresses the amplitude of the nucleon decays by a factor of $M^{2}$, so the nucleon stability experiments will put less stringent conditions on $M$. We shall be looking at specific models, extending the SM with renormalizable terms. (The effective operators would then correspond to dimension nine or twelve, instead of the dimension six operators considered in Refs.\cite{Hou:2005iu,Dong:2011rh}.)

In Section.\ref{sec:cat} we review the possible scalar leptoquarks and diquarks and their BNV interactions. We examine the constraints from nucleon stability on one of the interactions, considering couplings to first and third generation quarks. In Section.\ref{sec:washout} we analyse the effect such an interaction can have on baryogenesis, showing stringent limits apply for models which break the global $(B-L)$ symmetry. Finally, in Section.\ref{sec:colliders} we examine the phenomenology of this interaction at the LHC. Detection of such BNV interactions may then allow us to rule out certain baryogenesis mechanisms, even if their mass scale lies outside the reach of present day colliders. 

\section{Baryon Number Violating Scalars}
\label{sec:cat}
\subsection{A catalogue of models}
\begin{figure}[h]
	\begin{center}
	\includegraphics[width=200pt]{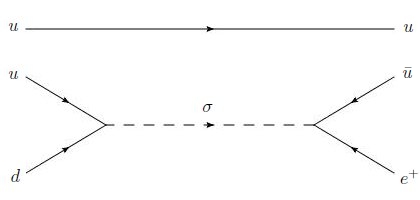}
	\end{center}
  	\caption{Proton decay, $p\to\pi^{0}e^{+}$, in a one exotic scalar extension with coupling to the first generation.}
	\label{fig:pion}
	\end{figure}

The models studied in this paper are those introduced in Ref.\cite{catalogue}, and were motivated by charge quantization. Here we use the same notation for the list of all possible scalar leptoquarks and diquarks, and their transformation under the SM gauge group $SU(3)_{C}\times SU(2)_{L} \times U(1)_{Y}$. The convention here is that the scalar fields have the same quantum numbers as the quark and lepton bilinears: 
	\begin{equation}	
	\label{eq:list}
	\begin{array}{l l l}
	\sigma_{1.1} & \sim \overline{Q_{L}}(f_{L})^{c} \sim \overline{u_{R}}(e_{R})^{c} \sim \overline{d_{R}}(\nu_{R})^{c} & \sim (\bar{3},1,2/3) \\
	\sigma_{1.2} & \sim \overline{Q_{L}}(f_{L})^{c} 							     & \sim (\bar{3},3,2/3) \\
	\sigma_{2}   & \sim \overline{Q_{L}}e_{R} \sim \overline{u_{R}}f_{L}					     & \sim (\bar{3},2,-7/3) \\
	\sigma_{3.1}   & \sim \overline{Q_{L}}(Q_{L})^{c} \sim \overline{u_{R}}(d_{R})^{c}                   & \sim (3,1,-2/3) \\
	\sigma_{3.2}   & \sim \overline{Q_{L}}(Q_{L})^{c}				             & \sim (3,3,-2/3) \\
	\sigma_{3.3}   & \sim \overline{Q_{L}}(Q_{L})^{c} \sim \overline{u_{R}}(d_{R})^{c}	             & \sim (\bar{6},1,-2/3)\\		
	\sigma_{3.4}   & \sim \overline{Q_{L}}(Q_{L})^{c}	                                     & \sim (\bar{6},3,-2/3)\\		
	\sigma_{4}   & \sim \overline{u_{R}}(\nu_{R})^{c}	                                     & \sim (\bar{3},1,-4/3)\\	
	\sigma_{5}   & \sim \overline{d_{R}}f_{L} \sim \overline{Q_{L}}\nu_{R}                         & \sim (\bar{3},2,-1/3)\\	
	\sigma_{6.1}   & \sim \overline{u_{R}}(u_{R})^{c}                          & \sim (3,1,-8/3)\\	
	\sigma_{6.2}   & \sim \overline{u_{R}}(u_{R})^{c}                          & \sim (\bar{6},1,-8/3)\\	
	\sigma_{7.1}   & \sim \overline{d_{R}}(d_{R})^{c}                          & \sim (3,1,4/3)\\	
	\sigma_{7.2}   & \sim \overline{d_{R}}(d_{R})^{c}                          & \sim (\bar{6},1,4/3)\\	
	\sigma_{8}   & \sim \overline{d_{R}}(e_{R})^{c}                          & \sim (\bar{3},1,8/3),\\	
	\end{array}
	\end{equation}
 where $\Psi^{c}\equiv\mathrm{C}\bar{\Psi}^{T}$, and C is the charge conjugation matrix. The standard model fermions and right-handed neutrinos transform in the usual way:
	\begin{equation}
	\begin{array}{l l l}
	f_{L} \sim (1,2,-1), & e_{R} \sim (1,1,-2), & \nu_{R} \sim (1,1,0), \\
	Q_{L} \sim (3,2,1/3), & u_{R} \sim (3,1,4/3), & d_{R} \sim (3,1,-2/3).
	\end{array}
	\end{equation}
Note the leptoquarks in Eq.(\ref{eq:list}) carry $B=-1/3$, and the diquarks $B=-2/3$. As we are breaking this symmetry, the following particles then carry identical quantum numbers:
	\begin{equation}
	\label{eq:pairs}
	\begin{array}{l}
	\sigma_{1.1} = \sigma_{3.1}^{c} \sim (\bar{3},1,2/3) \\ 
	\sigma_{1.2} = \sigma_{3.2}^{c} \sim (\bar{3},3,2/3) \\
	\sigma_{4} = \sigma_{7.1}^{c} \sim (\bar{3},1,-4/3) \\
	\sigma_{6.1} = \sigma_{8}^{c} \sim (3,1,-8/3).
	\end{array}
	\end{equation}
Such particles with both leptoquark and diquark couplings leads to baryon number violation. For example, taking the $\sigma \equiv \sigma_{1.1} = \sigma_{3.1}^{c} \sim (\bar{3},1,2/3)$ model the Lagrangian is extended to include the terms:
	\begin{equation}
	\label{eq:conjugate}
	\mathcal{L} \supset \lambda_{1.1}\overline{(e_{R})^{c}}\sigma u_{R} + \lambda_{3.1}\overline{(u_{R})}\sigma d_{R}^{c} + H.c.,
	\end{equation}
where $\lambda_{1.1}$ and $\lambda_{3.1}$ are dimensionless coupling constants, and generational and colour indices have been suppressed. If we take both Yukawa couplings to the SM fermions to be $\lambda\equiv\lambda_{1.1}\approx\lambda_{3.1}$, we obtain the following estimate of the proton decay rate on dimensional grounds (see fig.\ref{fig:pion})\cite{catalogue}:
	\begin{equation}
	\Gamma \sim \mathcal{O}\left(\frac{\lambda^{4}M_{p}^{5}}{M_{\sigma}^{4}}\right),
	\end{equation}
where $M_{p}$ is the proton mass. Given the bound on the partial lifetime for $p \to \pi^{0} e^{+}$, $\tau > 8.2\times10^{33}$ years \cite{:2009gd}, this translates to a limit on the mass of:
	\begin{equation}
	\label{eq:mass}
 	M_{\sigma} \gtrsim \lambda \times 10^{16} \mathrm{GeV}.
	\end{equation}

	\begin{figure}[h]
	\begin{center}
	\includegraphics[width=200pt]{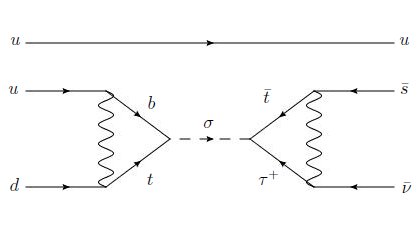}
	\end{center}
  	\caption{Proton decay, $p\to K^{+}\bar{\nu}$, in a one exotic scalar extension with coupling to the third generation.}
	\label{fig:kaon}
	\end{figure}

If we instead take the dominant coupling to be to the third generation, proton decay now proceeds through the diagram in Fig.\ref{fig:kaon}. The decay rate is estimated as (see appendix for details):
	\begin{align}
	\label{eq:kaon}
	\Gamma \sim & \frac{\lambda^{4}g_{w}^{8}M_{p}^{5}}{M_{\sigma}^{4}} \nonumber \\
	& \times \left( \frac{|V_{ub}||V_{td}||V_{ts}|M_{b}M_{\tau}}{M_{t}^{2}}\mathrm{ln}\left[\frac{M_{t}}{M_{b}}\right]\mathrm{ln}\left[\frac{M_{t}}{M_{\tau}}\right] \right)^{2},
	\end{align}
where $g_{w}$ is the weak coupling constant. Given the limit on the partial lifetime for $p \to K^{+}\bar{\nu}$, $\tau > 2.3\times10^{33}$years\cite{Kobayashi:2005pe}, this translates to a limit on the mass of:
	\begin{equation}
	\label{eq:3genmass}
	M_{\sigma} \gtrsim \lambda \times 10^{11}\mathrm{GeV}.
	\end{equation}
This corresponds to ``opening up" the effective operator
	\begin{equation}
	O^{(5)}_{ttbl}=(\overline{(b_{R})^{c}}t_{R})(\overline{(t_{R})^{c}}l_{R})
	\end{equation}
of Ref.\cite{Hou:2005iu} and is comparable to their limit on its coefficient $C^{(5)}_{ttbl} < 10^{-11}\mathrm{TeV}^{-2}$. Such BNV physics is then only accessible at the LHC if we assume a GIM-like cancellation mechanism to avoid these stringent bounds. We will not investigate the possibility of such a mechanism here. Instead we restrict our attention to the scalars which are not subject to this large mass bound, namely: $\sigma_{3.3},\sigma_{3.4},\sigma_{5},\sigma_{6.2},\sigma_{7.2}$.

Baryon number is then broken by the introduction of two scalars and an interaction between them. The possible scalar-scalar interactions, without the particles of Eq.(\ref{eq:pairs}), are listed below.
The $\Delta B=1$ list is:
	\begin{equation}
	\label{eq:listtwo}
		\begin{array}{l l l}	
		\sigma_{5},\rho		& \to & \sigma_{5}\sigma_{5}\sigma_{5}\rho \\
		\sigma_{5}^{a},\sigma_{5}^{b} & \to & \sigma_{5}^{a}\sigma_{5}^{a}\sigma_{5}^{b}\phi^{0}, \\
		\end{array}
	\end{equation}
where $\phi^{0}$ is the physical SM Higgs scalar, $\rho$ represents a new Higgs like scalar $\rho \sim (8,2,1)$, and $\sigma_{5}^{a},\sigma_{5}^{b}$ are the components of the $\sigma_{5}$ $SU(2)_{L}$ doublet. The $\Delta B=2$ list is:
	\begin{equation}
	\label{eq:listthree}
		\begin{array}{l l l}
		\sigma_{3},\sigma_{7} 	& \to & \sigma_{3.3}\sigma_{3.3}\sigma_{7.2} \\	
					& \to & \sigma_{3.4}\sigma_{3.4}\sigma_{7.2} \\			
		\sigma_{6},\sigma_{7} & \to & \sigma_{6.2}\sigma_{7.2}\sigma_{7.2}. \\
		\end{array}
	\end{equation}		
Note that the models of Eq.(\ref{eq:listtwo}) conserve ($B-L$) while the models of Eq.(\ref{eq:listthree}) break ($B-L$). This is an important distinction when it comes to consideration of the interaction's effects on baryogenesis. Breaking baryon number using these scalar-scalar interactions leads to less stringent bounds from nucleon stability. 

Considering this shorter list, we will study the $\sigma_{3.3}\sigma_{3.3}\sigma_{7.2}$ interaction in greater detail, due to its interesting phenomenology. First of all, it involves diquarks, which can be produced more favourably than leptoquarks in hadron colliders. Secondly, $\sigma_{3.3}$ couples to an up and down type quark. This means the possibility of top quarks in the final state, which is a clearer signal than generic jets. Finally, for simplicity, we examine the $SU(2)_{L}$ singlet, $\sigma_{3.3}$, rather than the $SU(2)_{L}$ triplet, $\sigma_{3.4}$.

The $\sigma$ particles in Eq.(\ref{eq:list}), are either scalar diquarks or scalar leptoquarks. The usual constraints, from low energy experiments, on the couplings of diquarks and leptoquarks to standard model fermions then apply. For example, from consideration of atomic parity violation of caesium, leptoquarks coupled to the right handed first generation fermions have a bound $M > \lambda \times 2$TeV \cite{PhysRevD.49.333}. For further constraints on the leptoquarks (in particular on products of different couplings) see Refs.\cite{PhysRevD.49.333,Davidson:1993qk,PhysRevD.81.095011,Dorsner:2011ai}. Products of diquark couplings are also constrained from limits on flavour changing neutral currents (FCNC) and meson-antimeson oscillations \cite{Giudice:2011ak}. These limits, however, do not apply in the limit where all but one of the couplings vanish.

There are also constraints on the masses of leptoquarks and diquarks from collider experiments. From searches of pair produced leptoquarks the best limits for the three generations are \cite{Aad:2011uv,Khachatryan:2010mp,Aad:2011uv,Khachatryan:2010mq,PhysRevLett.85.2056}:
\begin{equation}
\begin{array}{l l l }
M_{LQ} > & 384\mathrm{GeV} &\text{1st generation} \, (jjee)    \\
M_{LQ} > & 420\mathrm{GeV} &\text{2nd generation} \, (jj\mu\mu)  \\ 
M_{LQ} > & 148\mathrm{GeV} &\text{3rd generation} \, (bb\nu\nu),  \\
\end{array}
\end{equation}
where we have indicated the final state searched for with $j=$jet.

Diquarks can be produced singly at hadron colliders and then decay to two jets. It is then possible to exclude a given diquark given its coupling and mass from searches of dijet resonances \cite{PhysRevD.55.R5263,Chatrchyan:2011ns,Collaboration:2011fq}. Unlike the searches for pair produced leptoquarks, however, this diquark search is model dependent as the production occurs through the Yukawa coupling to the SM quarks.

As noted before, such baryon number violating scalar interactions can also be the mechanism for baryogenesis \cite{sigmabaryogenesis}. The $\sigma_{3.3}\sigma_{3.3}\sigma_{7.2}$ model, studied in the rest of the paper, was proposed as a baryogenesis mechanism by the authors of Ref.\cite{Babu:2006xc,Babu:2008rq}. The addition of a singlet scalar field, which decays out of equilibrium, allows the third Sakharov condition to be met, even with $\mathcal{O}$(TeV) masses for the coloured scalars. 

Alternatively, colour triplet scalars with both diquark couplings to the SM and leptoquark coupling to a SM quark and a right handed Majorana neutrino would also allow the third Sakharov condition to be met \cite{Gu:2011ff}. These papers also discuss the neutron-antineutron oscillations induced by such fields \cite{PhysRevLett.44.1316}.

Here we do not consider these particles as a baryogenesis mechanism, instead we obtain constraints from washout of baryogenesis.

\subsection{The Particular Model: $\sigma_{3.3}\sigma_{3.3}\sigma_{7.2}$}
In the remainder of this paper we shall focus on the $\sigma_{3.3}\sigma_{3.3}\sigma_{7.2}$ model. This model has the advantage of not containing any of the conjugate pairs of Eq.(\ref{eq:pairs}): recall that if the particle has both diquark and leptoquark couplings, the stringent limit on the mass of Eq.(\ref{eq:3genmass}) applies. In addition the interaction $\sigma_{3.3}\sigma_{3.3}\sigma_{7.2}$ can produce top quarks in the final state which would give an interesting signal at the LHC. Furthermore, the production of single diquarks through the Yukawa coupling is more favourable than leptoquarks at hadron colliders. The SM Lagrangian is extended to include the terms:
	\begin{equation}
	\mathcal{L} \supset \lambda_{3}\overline{(d_{R})^{c}}u_{R}\sigma_{3.3} +  \lambda_{7}\overline{(d_{R})^{c}}d_{R}\sigma_{7.2} + \mu \sigma_{3.3} \sigma_{3.3} \sigma_{7.2} + H.c.
	\end{equation}
where $\lambda_{3}$, $\lambda_{7}$ are dimensionless coupling constants and $\mu$ has dimension of mass. Generational and colour indices have been suppressed. The exotic fields transform in the following way under the SM gauge interactions:
	\begin{equation}
	\begin{array}{l l}
	\sigma_{3.3} \sim (\bar{6},1,-2/3), & \sigma_{7.2} \sim (\bar{6},1,4/3).
	\end{array}
	\end{equation}
We will primarily be interested in the $M_{\sigma_{7}} > 2M_{\sigma_{3}}$ hierarchy, especially for phenomenology at the LHC, which we study in Sec.\ref{sec:colliders}.

The couplings $\lambda_{3}$ and $\lambda_{7}$ are constrained from low energy experiments measuring FCNC and neutral meson oscillations. For example if the coupling to the third generation quarks $\lambda_{7}^{bb}=0.1$, the coupling to the second generation quarks is constrained to be $\lambda_{7}^{ss} < 10^{-3}$, for $M_{\sigma_{7}}=1$TeV or the $B^{0}_{s}-\overline{B^{0}_{s}}$ oscillation frequency would exceed the experimental value \cite{Giudice:2011ak,PhysRevD.77.011701,D'Ambrosio2001123}.

	\begin{figure}[h]
	\begin{center}
	\includegraphics[width=200pt]{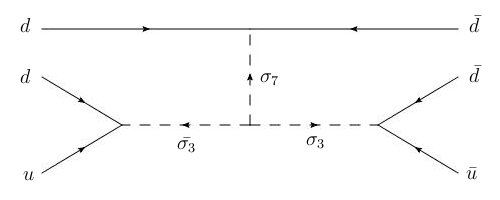}
	\end{center}
  	\caption{Tree level $n-\bar{n}$ oscillation for the $\sigma_{7}\sigma_{3}\sigma_{3}$ model with the scalars coupled to the first generation quarks.}
	\label{fig:n_osc}
	\end{figure}

	\begin{figure}[h]
	\begin{center}
	\includegraphics[width=250pt]{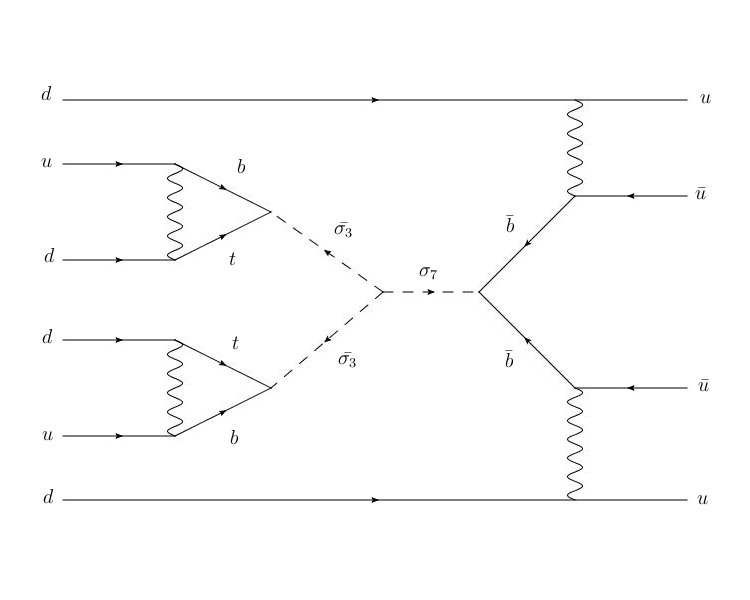}
	\end{center}
  	\caption{Two loop $nn \to \pi^{0}\pi^{0}$ decay for the $\sigma_{7}\sigma_{3}\sigma_{3}$ model with the scalars coupled to the third generation quarks.}
	\label{fig:nloop}
	\end{figure}	

	\begin{figure}[h]
	\begin{center}
	\includegraphics[width=250pt]{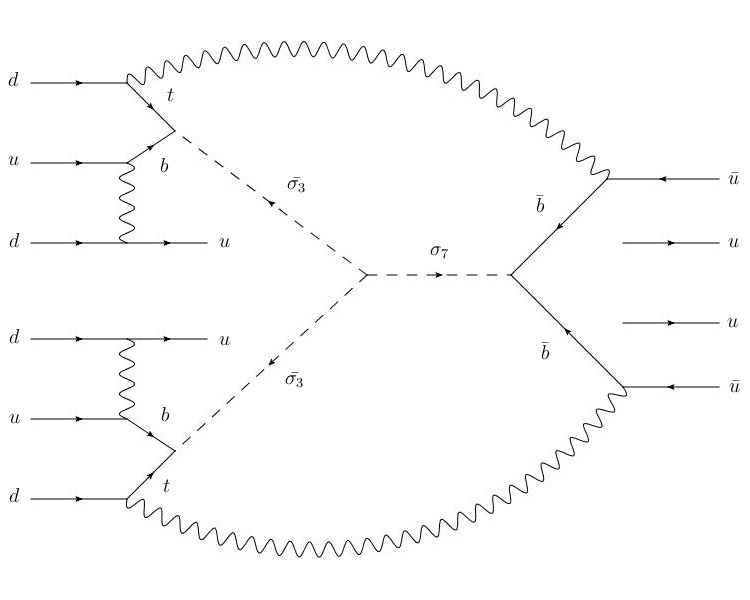}
	\end{center}
  	\caption{Two loop $nn \to \pi^{0}\pi^{0}$ decay for the $\sigma_{7}\sigma_{3}\sigma_{3}$ model with the scalars coupled to the third generation quarks.}
	\label{fig:mloop}
	\end{figure}

For couplings to the first generation quarks, there is also a stringent constraint from neutron-antineutron ($n-\bar{n}$) oscillations (see Fig.\ref{fig:n_osc}). Using dimensional analysis, the rate of oscillation is estimated as:
	\begin{equation}
	\Gamma(n-\bar{n}) \sim \mathcal{O}\left(\frac{\mu\lambda_{7}\lambda_{3}^{2}M_{n}^{6}}{M_{\sigma_{7}}^{2}M_{\sigma_{3}}^{4}}\right),
	\end{equation} 
where $M_{n}$ is the mass of the neutron. The lower limit from experiment for the oscillation time is currently around $\tau > 10^8$s \cite{PhysRevD.66.032004,springerlink:10.1007/BF01580321}. Taking $\lambda \equiv \lambda_{3} \approx \lambda_{7}$, $M_{\sigma} \equiv M_{\sigma_{3}} \sim M_{\sigma_{7}}$, and
$\mu \sim \lambda M_{\sigma}$, we obtain a bound:
	\begin{equation}
	\label{eq:low}
	M_{\sigma} \gtrsim \lambda^{4/5} \times 10^{6} \mathrm{GeV.}
	\end{equation}
A similar constraint from dinucleon decay has previously been calculated\cite{catalogue}.

We have set $\mu \sim \lambda M_{\sigma}$ as this ensures the two decay modes of $\sigma_{7}$ have similar branching ratio. In the limit $\mu \to 0$, the BNV aspect of this model is switched off and no constraints from nucleon decay or washout exist. A small $\mu$, however, makes the branching ratio for $\bar{\sigma_{7}} \to \sigma_{3}+\sigma_{3}$ insignificant. Detecting any BNV interaction at colliders then becomes even more unlikely. 

The constraint on $M_{\sigma}$, however, changes when the dominant coupling is taken to be to the third generation SM fermions. The most stringent constraint then comes from $nn \to \pi^{0}\pi^{0}$ (see Figs.\ref{fig:nloop},\ref{fig:mloop}), which has a partial lifetime $\tau > 3.4\times10^{30}$ years \cite{Berger1991227}. Evaluating the integrals around the loops, and then using dimensional analysis and taking $\mu \sim \lambda M_{\sigma}$, the double neutron decay rate is estimated to be:
	\begin{equation}
	\label{eq:double}
	\Gamma \sim \mathcal{O}\left(\frac{\lambda^{8}|V_{ub}|^{8}|V_{td}|^{4}g_{w}^{16}M_{n}^{23}}{M_{W}^{8}M_{t}^{4}M_{\sigma}^{10}}\right).
	\end{equation}
This translates into a limit on the mass of:
	\begin{equation}
	M_{\sigma} \gtrsim \lambda^{4/5}\times10\mathrm{GeV.}
	\end{equation}
The simplest $n-\bar{n}$ diagram now involves four loops, so we expect it to be even more suppressed. If the particles couple predominantly to the third generation, we see that nucleon stability constraints do not preclude BNV from occurring at LHC energies. We will soon see, however, that far more stringent constraints arise from washout of baryogenesis.

\begin{figure}[h]
	\begin{center}
	\includegraphics[width=200pt]{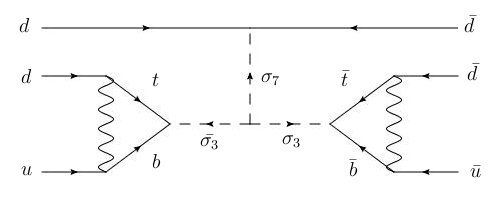}
	\end{center}
  	\caption{Two loop $n-\bar{n}$ oscillation for the $\sigma_{7}\sigma_{3}\sigma_{3}$ model with $\sigma_{7}$ coupled to the first and $\sigma_{3}$ to the third generation quarks.}
	\label{fig:inter}
	\end{figure}

Finally, let us mention an intermediate case where the $\sigma_{7}$ couples to the first generation and the $\sigma_{3}$ to the third. The limit on the mass again comes from $n-\bar{n}$ oscillations (see Fig.\ref{fig:inter}). The oscillation rate is estimated as:
	\begin{equation}
	\Gamma(n-\bar{n}) \sim \mathcal{O} \left( \mu\lambda_{7}\lambda_{3}^{2} \left(\frac{M_{b}}{M_{t}}\right)^{2} \frac{g_{w}^{4}|V_{td}|^{2}|V_{ub}|^{2}}{M_{\sigma_{7}}^2M_{\sigma_{3}}^4} M_{n}^{6} \right).
	\end{equation}
This translates as a limit on the mass of:
	\begin{equation}
	\label{eq:inter}
	M_{\sigma} \gtrsim \lambda^{4/5} \times 6 \mathrm{TeV}.
	\end{equation}

\section{Washout of Baryogenesis}
\label{sec:washout}
\subsection{High Temperature Baryogenesis}
If baryogenesis were to occur before the temperature, $T$, of the universe reached the mass scale of the $\sigma$ particles, as would be the case in some high temperature leptogenesis scenario, the creation and decay of the $\sigma$ particles in a $\Delta(B-L) \neq 0$ sequence can erase the matter-antimatter asymmetry. For example, in the $M_{\sigma_{7}} > 2M_{\sigma_{3}}$ case such a sequence could proceed on-shell in the following way:
	\begin{align}
	d+d \to \overline{\sigma_{7}} \\
	\overline{\sigma_{7}} \to \sigma_{3}+\sigma_{3} \\
	\sigma_{3} \to \bar{u}+\bar{d}.
	\end{align}
If the rate of one of these steps, $\Gamma$, is less than the expansion rate of the universe, $H$, the $\Delta(B-L) \neq 0$ process will not be occurring rapidly enough for washout \cite{PhysRevLett.45.2074,PhysRevD.22.2953,PhysRevD.22.2977}. If we again take the $\lambda \equiv \lambda_{3} \approx \lambda_{7}$, $\mu \sim \lambda M_{\sigma}$ case we see to avoid washout we require:
	\begin{equation}
	\label{eq:onshell}
	\Gamma \sim \lambda^{2}M_{\sigma}\frac{M_{\sigma}}{T} < H \sim \frac{g^{1/2}T^{2}}{M_{Pl}},
	\end{equation}
where $g$ specifies the degrees of freedom, $M_{Pl}$ is the Planck mass, and we have included a Lorentz factor for the typical rate. Remembering that this is required to hold for all $T>M_{\sigma}$ (below which the initial inverse decay will be Boltzmann suppressed), this translates into a washout avoidance condition on $\lambda$:
	\begin{equation}
	\label{eq:hight}
	\lambda \lesssim \left(\frac{g^{1/2}M_{\sigma}}{M_{Pl}}\right)^{1/2}.
	\end{equation}
This is a more stringent constraint than Eq.(\ref{eq:low}) for $\mathcal{O}\mathrm{(TeV)}$ masses. So washout of high temperature leptogenesis could take place with the addition of such fields, even with couplings only to the first generation (which are the most constrained from nucleon stability). 

The addition of such fields with couplings greater than the above constraint, would mean any existing asymmetry in the quark sector above the mass scale of the $\sigma$ particles will be removed by their BNV interactions.

\subsection{Low Temperature Baryogenesis}
Now we examine the scenario where baryogenesis occurs at a temperature below the mass scale of the $\sigma$ particles, which may be the case e.g. in electroweak baryogenesis. The inverse decay $dd\to\bar{\sigma_{7}}$ is now Boltzmann suppressed:
	\begin{equation}
	\Gamma \sim \lambda^{2}M_{\sigma}\left(\frac{M_{\sigma}}{{T}}\right)^{3/2} \mathrm{exp}\left(\frac{-M_{\sigma}}{{T}}\right).
	\end{equation} 
To avoid washout we require this rate to be less than $H$ at the temperature of baryogenesis, $T_{b}$, giving the bound:
	\begin{equation}
	\label{eq:lowtavoid}
	\lambda \lesssim \left( \mathrm{exp}\left(\frac{M_{\sigma}}{{T_{b}}}\right)\frac{g^{1/2}T_{b}^{7/2}}{M_{Pl}M_{\sigma}^{5/2}} \right )^{1/2}.
	\end{equation}
Setting $M_{\sigma}=1$TeV and $T_b=100$GeV, we obtain a bound of:
	\begin{equation}
	\label{eq:lowtavoidnum}
	\lambda \lesssim 10^{-7}.
	\end{equation}
Due to there being only one order of magnitude difference between $M_{\sigma}$ and $T_{b}$, the inverse decay is not yet sufficiently suppressed for the on-shell sequence to be insignificant, and thus the stringent bound on $\lambda$, Eq.(\ref{eq:lowtavoid}), applies.

	\begin{figure}[h]
	\begin{center}
	\includegraphics[width=200pt]{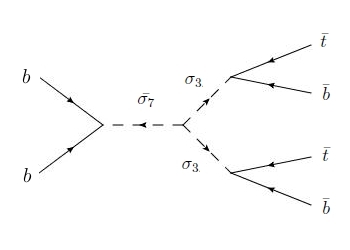}
	\end{center}
  	\caption{BNV scattering for the $\sigma_{7}\sigma_{3}\sigma_{3}$ model}
	\label{fig:wash}
	\end{figure}

If $M_{\sigma}$ were increased substantially (to $\approx4.5$TeV), the inverse decay does become sufficiently suppressed. The off-shell BNV scattering process depicted in Fig.\ref{fig:wash} then becomes dominant, with rate:
	\begin{equation}
	\Gamma \sim \frac{\mu^{2}\lambda^{6}T^{11}}{M_{\sigma}^{12}}.
	\end{equation}
When compared to $H$, and again taking $\mu \sim \lambda^{2}M_{\sigma}$, this translates into the washout avoidance condition:
	\begin{equation}	
	\label{eq:avoidance}
	\lambda \lesssim \left(\frac{g^{1/2}M_{\sigma}^{10}}{T_{b}^{9}M_{Pl}}\right)^{1/8},
	\end{equation}
which is more stringent for $T_{b}=100$GeV than the constraint from Eq.(\ref{eq:lowtavoid}) only for $M_{\sigma}\gtrsim4.5$TeV. For these larger masses, we can have $\lambda\sim0.1-1$ while still satisfying constraints from washout avoidance and nucleon stability.

For lower masses $M_{\sigma} \lesssim 4.5$TeV, the couplings of this $(B-L)$ violating interaction are tightly constrained by the washout avoidance condition, no matter to which generation quarks the scalars couple. A stronger coupling would mean baryogenesis would be washed out if it occurred at a temperature $T_{b}>100$GeV. As we are almost certain it occurred at a temperature of at least 100GeV, we take the constraint of Eq.(\ref{eq:lowtavoidnum}) as our starting point for our examination of this model at the LHC for masses below 4.5TeV. We will return to the high mass case, $M_{\sigma} \gtrsim 4.5$TeV, at the end of the next section. 

Finally we point out that if $\lambda$ lies between the high and low temperature washout constraints, detection of such interactions could be used to rule out certain high temperature baryogenesis mechanisms, without having to probe their possibly high energy scales $\sim 10^{16}$GeV directly.

\section{Collider Searches}
\label{sec:colliders}
The production of the exotic scalars can take place either through the Yukawa coupling with the quark fields or by gluon-gluon fusion. Let us first examine both possibilities, in the case where the heaviest diquark mass is $M_{\sigma_{7}} \lesssim 4.5$TeV, in light of the stringent washout avoidance constraint in Eq.(\ref{eq:lowtavoidnum}). 

In the limit of $\mu\to0$, we effectively turn off the BNV interaction and return to the standard diquark phenomenology. The Yukawa coupling to the first generation may then be large enough for the $qq \to \sigma \to jj$ signal to be significant. Such diquark production has been studied in \cite{Han:2009ya,PhysRevD.77.011701,Gogoladze2010233}. Experimental limits on such dijet resonances have already been set \cite{PhysRevD.55.R5263,Chatrchyan:2011ns,Collaboration:2011fq}.

\begin{figure}[h]
	\begin{center}
	\includegraphics[width=200pt]{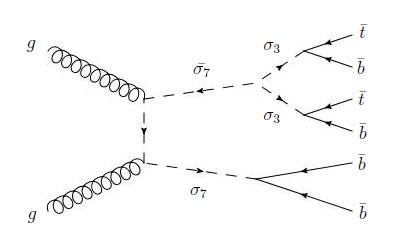}
	\end{center}
  	\caption{Pair production of $\sigma_{7}\bar{\sigma_{7}}$ through gluon-gluon fusion. The $\bar{\sigma_{7}}$ decays into $\sigma_{3}\sigma_{3}$ resulting in a $\bar{t}\bar{b} \; \bar{t}\bar{b} \; \bar{b}\bar{b}$ final state. The $bb \; tb \; tb$ final state from the $\sigma_{7}\bar{\sigma_{7}}$ pair production would occur in the same amount.}
	\label{fig:gluons}
	\end{figure}

If $\mu \sim \lambda M_{\sigma}$, the washout avoidance condition implies a lifetime $c\tau > 10^{-5}$m, so the observation of displaced vertices may occur. The scalars would be pair produced through gluon-gluon fusion (see Fig.\ref{fig:gluons}), which depends only on the colour charge. For a unity branching ratio into two jets, the LHC at $\sqrt{s}=14$TeV with $\mathcal{L}=100\mathrm{fb}^{-1}$ of data, can discover such pair produced colour sextet scalars with masses below 1050GeV \cite{Richardson:2011df,Chen:2008hh}. If we assume the hierarchy $M_{\sigma_{7}}>2M_{\sigma_{3}}$, we must only concern ourselves with the branching ratio for $\bar{\sigma_{7}} \to \sigma_{3}+\sigma_{3}$ and to which generation quarks the $\sigma$ particles decay. If the coupling was to the first two generations, it would not be possible to extract the existence of a BNV process, due to the generic nature of the jets. 

Reference \cite{DelNobile:2009st} provides a plot of the pair production cross section of colour sextet scalars as a function of the particle mass. Pair production of the scalars declines from $\approx6$pb for $M=500$GeV to $\approx60$fb for $M=1$TeV at $\sqrt{s}=14$TeV. Taking $M_{\sigma_{7}}>2M_{\sigma_{3}}$, branching ratio $r$ for $\bar{\sigma_{7}} \to \sigma_{3}\sigma_{3}$, and coupling of the scalars to the third generation quarks we have the final states of pair produced $\sigma_{7}\bar{\sigma_{7}}$ in Table.I (also see Fig.\ref{fig:gluons} for an example of the decay chain).

\begin{table}[h!]
\begin{center}
    \begin{tabular}{ | l | p{3cm} | }
    \hline
    Final state & Fraction of total \\ \hline

    $bb\bar{b}\bar{b}$ & $(1-r)^{2}$ \\ \hline

    $bb \; tb \; tb$ & $r(1-r)$ \\ \hline

    $\bar{t}\bar{b} \; \bar{t}\bar{b} \; \bar{b}\bar{b}$  & $r(1-r)$    \\ \hline

    $\bar{t}\bar{b} \; \bar{t}\bar{b} \; tb \; tb$  & $r^{2}$ \\
    \hline
    \end{tabular}
\end{center}
\caption{Final states from pair production of $\sigma_{7}\bar{\sigma_{7}}$ given branching ratio $r$ for $\bar{\sigma_{7}} \to \sigma_{3}\sigma_{3}$, and coupling of the scalars to the third generation quarks.}
\end{table}

One could then utilise the semi leptonic decay to distinguish top from anti-top. This would reduce the signal by a factor of (2/9) for every top quark required to decay to a lepton. But we are still left with many $b$ jets, the charge of which is even more difficult to extract. Furthermore due to the initial $B=0$ $\sigma_{7}\bar{\sigma_{7}}$ state, there are the same number of quark and antiquark jets over a many event average. Such a BNV signal is therefore unlikely to be detected conclusively at the LHC.

Now to the case $\lambda\to0$ and $\mu \gg \lambda M_{\sigma}$. Here the phenomenology does not change much, except for large enough $\mu$, the decay width $\overline{\sigma_{7}} \to \sigma_{3}+\sigma_{3}$ becomes large and $\sigma_{7}$ may then not decay at a displaced vertex. Given the mass hierarchy we have chosen, $M_{\sigma_{7}}>2M_{\sigma_{3}}$, $\sigma_{3}$ will still decay preferably to two jets rather than through the virtual $\sigma_{7}$ propagator.

Finally we examine the case where the heaviest diquark mass is large: $M_{\sigma_{7}} \gtrsim 4.5$TeV. The couplings are then far less constrained from washout and there may be significant production $dd \to \bar{\sigma_{7}}$, as well as a significant branching fraction $\bar{\sigma_{7}} \to \sigma_{3}+\sigma_{3}$.  For purposes of illustration we choose values $\lambda_{7}=1$ and $r(\bar{\sigma_{7}} \to \sigma_{3}\sigma_{3})=0.5$. Furthermore we assume $\sigma_{3}$ decays to the third generation quarks. The choice $\lambda_{7}=1$ is consistent with the limit from washout: Eq.(\ref{eq:avoidance}). It saturates the nucleon stability limit, Eq.(\ref{eq:inter}), to the level of precision with which that limit was calculated.

\begin{figure}[h]
	\begin{center}
	\includegraphics[width=200pt]{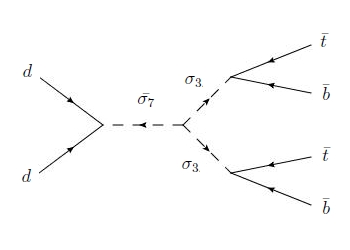}
	\end{center}
  	\caption{BNV at the LHC $dd \to \bar{\sigma_{7}} \to \sigma_{3}+\sigma_{3}$.}
	\label{fig:lhcbnv}
	\end{figure}

Using the Breit-Wigner approximation for scattering through resonances and the CT10 parton distribution function \cite{Lai:2010vv}, we obtain the cross sections for $\bar{t}\bar{b}\bar{t}\bar{b}$ and $jj$ final states in Table.II (also see Fig.\ref{fig:lhcbnv}).

\begin{table}[h!]
\label{tab:highmass}
\begin{center}
	\begin{tabular}{ | l || l | l | l | }
			  \hline
			  $M_{\sigma7}$ (TeV) & 4.75 & 5.0 & 5.25 \\ \hline \hline
			  $\sigma(\sqrt{s}=14$TeV) (pb) & 0.19 & 0.17 & 0.16  \\ \hline
			  $\sigma(\sqrt{s}=7$TeV) (pb) & 0.04 & 0.04 & 0.03  \\ \hline

	\end{tabular}
\end{center}
\caption{Final states $\sigma(\bar{t}\bar{b}\bar{t}\bar{b})$ = $\sigma(jj)$ from $\bar{\sigma_{7}}$ production for $pp$ collisions at $\sqrt{s}=14$TeV and $\sqrt{s}=7$TeV. The Yukawa coupling to the quarks is $\lambda_{7}=1$, and branching ratio $\bar{\sigma_{7}} \to \sigma_{3}\sigma_{3}$ equal to 0.5.}
\end{table}

Given that the limit on dijet resonances at $M=4.0$TeV at $\sqrt{s}=7$TeV is 0.005pb \cite{Chatrchyan:2011ns,Collaboration:2011fq}, such a high coupling $\lambda$ may very quickly be ruled out once more data is released (the experimental papers do not supply any limits for dijet resonances above 4.1TeV, we can imagine such a large dijet excess at 4.75-5.25TeV presumably would have already been pointed out if it existed). This scenario should therefore be taken as an illustration of the BNV signals accessible once the centre-of-mass energy is increased to $\sqrt{s}=14$TeV. Such a clear signal, however, relies on careful selection of the parameters and should therefore be taken as a best case scenario.

\section{Discussion}
Let us summarize the overall phenomenological picture for such BNV models. First the low mass regime for the heaviest scalar: $M_{\sigma_{7}} \lesssim 4.5$TeV. For the diquark interactions of Eq.(\ref{eq:listthree}), $(B-L)$ is broken so for such a low mass, one of the couplings must be small for washout of baryogenesis to be avoided. This forces us to rely on pair production for any sizable BNV process to be occurring at the LHC, making detection of the BNV interaction very difficult. Even without a clear BNV signal, multiple scalars may be discovered through the four jet pair production signal, or in the case $\mu\to0$, the dijet signal if $\lambda$ is large enough. 

Once the properties of the scalars begin to emerge, one would infer the possibility of a gauge-invariant cubic interaction between the recently discovered scalars. Thus the possibility of BNV in this sector of particles would be hinted at. The precision study of such an interaction would be left to further experimental work.

If the heaviest scalar mass is large, $M_{\sigma_{7}} \gtrsim 4.5$TeV, the washout condition is much weaker and a clearer BNV signal may occur albeit for a very specific choice of parameters. If such a choice is not realised we are again left with signals of multiple scalars. For masses greater than around $1$TeV, only the dijet signal governed by the Yukawa coupling to first generation quarks is accessible at the LHC, pair production being too small for such masses.

If we instead examine the $(B-L)$ conserving interactions of Eq.(\ref{eq:listtwo}) we are left with leptoquark instead of diquark scalars. The washout constraint then does not apply, but being leptoquarks, these again would not be produced singly in large numbers at hadron colliders (pair production through gluon-gluon fusion is still possible). That is, even in an optimistic scenario with EM strength Yukawa coupling, the production mechanisms $g+d \to d^{*} \to \bar{\sigma_{5b}} + e^{-}$ and $pp \to \gamma + d + X \to e^{+}e^{-}+d + X\to e^{-} + \bar{\sigma_{5b}} + X$, only result in a cross section of 3.4fb\cite{PhysRevD.57.1715} and 10fb\cite{Ohnemus:1994xf,PhysRevD.58.095001} respectively for $M_{\sigma}=1$TeV at $\sqrt{s}=14$TeV at the LHC. But due to the nucleon stability constraints this coupling is expected to be small. So we expect BNV leptoquarks to only be pair produced in large numbers in a hadron collider (not produced on their own). This again results in a $B=0$ initial state, making detection of the BNV interaction difficult. 

\section{Conclusion}
We have examined an extension of the standard model involving diquarks with renormalizable terms. We have shown that the nucleon stability constraints are greatly weakened if the exotic scalars are coupled to the third generation quarks. The diquark-diquark interaction violates ($B-L$), so for a mass of the heaviest diquark; $M_{\sigma}\lesssim4.5$TeV, the most stringent constraint now comes from washout of baryogenesis. For such a choice of mass we are left with very stringent constraints on at least one of the couplings no matter to which generation quarks the scalars couple. Although the new scalars can be produced at the LHC, these stringent coupling constraints greatly inhibit any clear identification of such BNV interactions at the LHC. If the heaviest diquark has a mass higher than 4.5TeV, the most stringent constraint again comes from nucleon stability. If the scalar coupling to the SM quarks then involves the higher generations, the couplings may be quite large, allowing for certain BNV processes to occur rapidly at LHC energies. In either case, even if no BNV interaction is detected conclusively, the exotic particles would show up in other ways in the experiments. The discovery of multiple scalars would be interesting in itself, and once an overall picture emerged, the possibility of BNV in this sector of particles could be postulated. 

\section*{Acknowledgements}
IB was supported by the Commonwealth of Australia, and NFB and RRV by the Australian Research Council. IB would like to thank A.Galea and N.Setzer for helpful discussions. 

\section*{Appendix}
To estimate the rate of the process in Fig.\ref{fig:kaon} we first evaluate the contribution from one of the fermion lines forming a loop with the $W$ boson. Using relations for the spinors such as $u = C\bar{v}^{T}$ (see appendix G.4 in \cite{Dreiner:2008tw}), and ignoring the external momenta we find the Feynman amplitude for such a loop:
	\begin{align}
	\bar{v}\int\frac{\mathrm{d}^{4}k}{(2\pi)^{4}}\frac{g^{\mu\nu}}{k^{2}-m_{W}^{2}}g_{w}V_{ub}\gamma^{\nu}\mathrm{R}\left(\frac{\slashed k+m_{b}}{k^{2}-m_{b}^{2}}\right)
	\nonumber	\\
	\times	\lambda_{3}\mathrm{R}\left(\frac{\slashed k+m_{t}}{k^{2}-m_{t}^{2}}\right)g_{w}V_{td}\gamma^{\mu}\mathrm{L}u,
	\end{align}
where L(R), is the left (right) chiral projection operator. This can be simplified to:
	\begin{align}
	(4 &\lambda_{3}V_{td}V_{ub}g_{w}^{2}m_{b}m_{t})  \times (\bar{v}\mathrm{L}u) \nonumber  \\
		& \times \left(\int\frac{\mathrm{d}^{4}k}{(2\pi)^{4}}\frac{1}{k^{2}-m_{w}^{2}}\frac{1}{k^{2}-m_{b}^{2}}\frac{1}{k^{2}-m_{t}^{2}}\right).
	\end{align}
Taking the leading term of the integral:
	\begin{equation}
	\frac{1}{8\pi^{2}m_{t}^{2}}\mathrm{ln}\left(\frac{m_{t}}{m_{b}}\right),
	\end{equation}
we see this fermion line contributes a factor to the overall amplitude of:
	\begin{equation}
	\frac{1}{2\pi^{2}}\frac{m_{b}}{m_{t}}\mathrm{ln}\left(\frac{m_{t}}{m_{b}}\right)\bar{v}\mathrm{L}u.
	\end{equation}
Then ignoring the external momenta, we find the dependence on the propagator masses for the overall diagram. The matrix element is simply estimated to be the mass of the neutron, with exponent set to restore the correct dimension of mass to the decay rate. This yields the rate of Eq.(\ref{eq:kaon}).

To estimate the rate of the process in Fig.\ref{fig:mloop} we note the integral around the loops has the form:
	\begin{align}
	\label{eq:second}
	\int & \frac{\mathrm{d}^{4}a}{(2\pi)^{4}} \Big ( \frac{1}{a^{2}-m_{w}^{2}}\frac{1}{a^{2}-m_{t}^{2}}\frac{1}{a^{2}-m_{b}^{2}}\frac{1}{a^{2}-m_{\sigma_{3}}^{2}} \nonumber \\
	& \times \Big \{ \int\frac{\mathrm{d}^{4}k}{(2\pi)^{4}}\frac{1}{k^{2}-m_{W}^{2}}\frac{1}{k^{2}-m_{t}^{2}}\frac{1}{k^{2}-m_{b}^{2}} \nonumber \\
	& \quad \quad \times \frac{1}{k^{2}-m_{\sigma_{3}}^{2}}\frac{1}{(k+a)^{2}-m_{\sigma_{7}}^{2}} \Big \} \Big ). 
	\end{align}
Focusing first on the integral over $k$, and introducing the Feynman parameters $x_{i}$, we rewrite this integral in the following form:
	\begin{align}
	4!\int_{0}^{1}&\int_{0}^{1-x_{1}}\int_{0}^{1-x_{2}-x_{1}}\int_{0}^{1-x_{3}-x_{2}-x{1}}\mathrm{d}x_{1}\mathrm{d}x_{2}\mathrm{d}x_{3}\mathrm{d}x_{4} \nonumber  \\
	& \times \int\mathrm{d}^{4}k \Big \{(k+ax_{1})^2+a^{2}x_{1}(1-x_{1}) \nonumber \\
	& \quad \quad \quad {} - (1-x_{1}-x_{2}-x_{3}-x_{4})m_{t}^{2} \nonumber \\
	& \quad \quad \quad {} - x_{1}m_{\sigma_{7}}^{2}-x_{2}m_{W}^{2}-x_{3}m_{\sigma_{3}}^{2}-x_{4}m_{b}^{2}+i\epsilon \Big \}^{-5}.
	\end{align}
(We have reintroduced the $i\epsilon$, $\epsilon > 0$ term in the denominator of the propagators.) After changing variable to $l = k+ax_{1}$ this becomes a standard integral which evaluates to:
	\begin{align}
	\frac{-2i}{(4\pi)^{2}}\int_{0}^{1}&\int_{0}^{1-x_{1}}\int_{0}^{1-x_{2}-x_{1}}\int_{0}^{1-x_{3}-x_{2}-x{1}}\mathrm{d}x_{1}\mathrm{d}x_{2}\mathrm{d}x_{3}\mathrm{d}x_{4} \nonumber \\
	& \times \Big \{ a^{2}x_{1}(1-x_{1})-(1-x_{1}-x_{2}-x_{3}-x_{4})m_{t}^{2} \nonumber \\
	& \quad {}  - x_{1}m_{\sigma_{7}}^{2}-x_{2}m_{W}^{2}-x_{3}m_{\sigma_{3}}^{2}-x_{4}m_{b}^{2}+i\epsilon \Big \}^{-3}.
	\end{align}
Now writing $a^{2}=a_{0}^{2}-a_{i}^{2}$, the denominator is zero for:
	\begin{align}
	a_{0} = & \pm \Big \{ \frac{1}{x_{1}(1-x_{1})}[m_{t}^{2}(1-x_{1}-x_{2}-x_{3}-x_{4}) \nonumber \\
	&	+x_{1}m_{\sigma_{7}}^{2}+x_{2}m_{W}^{2}+x_{3}m_{\sigma_{3}}^{2}+x_{4}m_{b}^{2}]+a_{i}^{2} \Big \}^{1/2} \mp i\epsilon'. \\ \nonumber
	\end{align}
This means any singularities occur in the second or fourth quadrant on the $a_{0}$ plane, allowing us to Wick rotate so that $a_{E0} = ia_{0}$ and $a_{E0}^{2} = -a^{2}$. 

Equation (\ref{eq:second}) now becomes:
\begin{align} 
\label{eq:third}
\frac{-2}{(4\pi)^{2}}  \int \frac{\mathrm{d}^{4}a_{E}}{(2\pi)^{4}} \Big(  \frac{1}{a_{E}^{2}+m_{w}^{2}} \frac{1}{a_{E}^{2}+m_{t}^{2}} \frac{1}{a_{E}^{2}+m_{b}^{2}} \frac{1}{a_{E}+m_{\sigma_{3}}^{2}} \nonumber \\
 \times \int_{0}^{1}\int_{0}^{1-x_{1}}\int_{0}^{1-x_{2}-x_{1}}\int_{0}^{1-x_{3}-x_{2}-x{1}}\mathrm{d}x_{1}\mathrm{d}x_{2}\mathrm{d}x_{3}\mathrm{d}x_{4}	 \nonumber \\
	\times \{ a_{E}^{2}x_{1}(1-x_{1})+(1-x_{1}-x_{2}-x_{3}-x_{4})m_{t}^{2} \nonumber \\
	\quad {}  + x_{1}m_{\sigma_{7}}^{2}+x_{2}m_{W}^{2}+x_{3}m_{\sigma_{3}}^{2}+x_{4}m_{b}^{2} \}^{-3} \Big).
\end{align}
Now this is in the form $\int \mathrm{d}a f(a)g(a)$, where $g(a)$ represents the integral over the Feynman parameters, with $f(a),g(a)$ positive definite. The integral over the Feynman parameters, $g(a)$, is maximal for $a=0$. If $g(a) \le A$ for all $a$, and $f(a),g(a)$ are positive definite  $\int \mathrm{d}a f(a)g(a) \leq \int \mathrm{d}a f(a)A$. Applying this to Eq.(\ref{eq:third}), we find the leading term of this integral to be:
	\begin{equation}
	\left(\frac{1}{m_{\sigma_{3}}^{2}m_{t}^{2}}\right)\left(\frac{1}{m_{t}^{2}m_{\sigma_{3}}^{2}m_{\sigma_{7}}^{2}}\right),
	\end{equation}
where the second bracket is the leading term of $g(a=0)$, and the first the remaining integral $\int \mathrm{d}a f(a)$, and we have ignored factors of $i$, $\pi$, and logarithms of $\mathcal{O}(1)$. Using a similar technique of dimensional analysis as outlined before we again obtain the rate of Eq.(\ref{eq:double}).

\end{document}